\documentclass[%
 aip,
 amsmath,amssymb,
 reprint,%
]{revtex4-1}

\usepackage{graphicx}% Include figure files
\usepackage{dcolumn}% Align table columns on decimal point
\usepackage{bm}% bold math
\usepackage[utf8]{inputenc}
\usepackage[T1]{fontenc}
\usepackage{mathptmx}
\usepackage{xcolor}
\begin{document}

\preprint{AIP/123-QED}

\title{Large-area microwire MoSi single-photon detectors 
at 1550 nm wavelength}
% Force line breaks with \\

\author{I. Charaev}
 \email{charaev@mit.edu.}
 \thanks{Both authors contributed equally to this work.}
\affiliation{ 
Massachusetts Institute of Technology, 50 Vassar Street, Cambridge, MA 02139, USA%\\This line break forced with \textbackslash\textbackslash
}%
\author{Y. Morimoto}
 \thanks{Both authors contributed equally to this work.}
\affiliation{ 
Massachusetts Institute of Technology, 50 Vassar Street, Cambridge, MA 02139, USA%\\This line break forced with \textbackslash\textbackslash
}%
\author{A. Dane}
\affiliation{ 
Massachusetts Institute of Technology, 50 Vassar Street, Cambridge, MA 02139, USA%\\This line break forced with \textbackslash\textbackslash
}%
\author{A. Agarwal}
\affiliation{ 
Massachusetts Institute of Technology, 50 Vassar Street, Cambridge, MA 02139, USA%\\This line break forced with \textbackslash\textbackslash
}%
\author{M. Colangelo}
\affiliation{ 
Massachusetts Institute of Technology, 50 Vassar Street, Cambridge, MA 02139, USA%\\This line break forced with \textbackslash\textbackslash
}%
\author{K. K. Berggren}
\affiliation{ 
Massachusetts Institute of Technology, 50 Vassar Street, Cambridge, MA 02139, USA%\\This line break forced with \textbackslash\textbackslash
}%
\date{\today}% It is always \today, today,

\begin{abstract}
We demonstrate saturated internal detection efficiency at 1550 nm wavelengths for meander-shaped superconducting nanowire single-photon detectors made of 3~nm thick MoSi films with widths of 1 and 3 $\mu$m, and active areas up to 400 by 400 $\mu$m$^2$. Despite hairpin turns and a large number of squares (up to 10$^4$) in the device, the dark count rate was measured to be $\sim$10$^3$ cps at 99\% of the switching current. This value is about two orders of magnitude lower than results reported recently for short MoSi devices with shunt resistors. We also found that 5~nm thick MoSi detectors with the same geometry were insensitive to single near-infrared photons, which may be associated with different levels of suppression of the superconducting order parameter. However, our results obtained on 3~nm thick MoSi devices are in a good agreement with predictions in the frame of a kinetic-equation approach. 
\end{abstract}

\maketitle

The superconducting nanowire single-photon detector (SNSPD) is an ideal detector for a broad range of applications such as space communication \cite{Brian2019}, LIDAR \cite{Li2017} and quantum key distribution \cite{Wei2019}. Following the first experimental demonstration of SNSPDs \cite{Goltsman2001}, several models have been proposed to explain the detection mechanism in nanowires \cite{Engel2015}. The deterministic hard-core and diffusive hot-spot models predict the threshold for single-photon detection for a given material and cross-section. It was also commonly accepted that the sensitivity is reduced with an increase in the dimensions of the nanowire (thickness, $d$ and/or width, $W$); however, a recently proposed theoretical model and experimental result involving single-photon response in micro-scale wires are in contradiction with previous understanding \cite{Korneeva2018}. According to the theory by D. Vodolazov\cite{Vod2017}, the detection of infrared photons in the narrow superconducting wires ($W \ll \Lambda$, where $\Lambda$ is the Pearl length) occurs once the ratio of critical current density, $j_C$ to the depairing current density, $j_C^{\rm dep}$ exceeds 0.7 at an operating temperature $T \approx T_C/2$ ($T_C$ - the critical temperature) independent of the cross-section of the structure. Experimentally, the single-photon regime of detection has been demonstrated on NbN \cite{Korneeva2018} and MoSi \cite{KorMoSi2018} with wires widths ranging from 0.5 up to 5 $\mu$m. 

Although the saturation of the internal detection efficiency for wide wires was not achieved at telecom wavelengths, which are of special interest, these results\cite{Korneeva2018, KorMoSi2018} indicated several practical advantages. First, wider nanowires make optical coupling and biasing easier than in traditional nanowire-based detectors. Second, microstructures can be patterned by photolithography which can be less expensive and less time consuming than electron or ion beam techniques. Also, the fabrication of large-area detectors by electron beam is a challenge due to requirements for long-term stability of electron beam parameters and long-term suppression of external acoustic, mechanical, and electro-magnetic interference. These drawbacks could be overcome by using a photolithographic technique. Once the photolithographic process is optimized in terms of the quality of the microstructure edges, the noise and electrical characteristics should be comparable with devices made by electron-beam lithography. This is a great help in scaling up the active area of detectors which is, for example, required for dark-matter detection by superconducting nanowires \cite{Hochberg2019, Baryakhtar2018}.

The film uniformity remains a decisive factor in scaling up the detectors. As noted by Gaudio and co-authors \cite{Gaudio2014}, the experimental switching current is observed to be continuously reduced with the increase of the length of the nanowire for NbN structures. Some of the possible reasons of such behavior are intrinsic constrictions due to polycrystalline films and variation of the superconducting gap across the film. 

As alternative materials for SNSPDs, amorphous WSi \cite{Marsili2013} and MoSi \cite{Verma2015, Caloz2017} films have been proposed and demonstrated in the last decade. Due to their amorphous microstructure, devices fabricated from these materials appear to suffer from fewer constrictions. SNSPDs made from either of these materials have shown excellent detection performance in the near infrared\cite{Li2016, Caloz2018}. 

In this work, we report on the fabrication and testing of large-area MoSi microwire single-photon detectors that exhibit saturation of the internal detection efficiency at 1550 nm wavelength. The morphological structure of thin MoSi films has been studied by transmission electron microscopy (TEM) analysis. We compare our obtained experimental results with previous reports and theoretical predictions.

The fabrication process begins with magnetron sputtering of MoSi films onto silicon substrates with a 300~nm thermally oxidized SiO$_2$ layer. Films were co-sputtered from Mo (99.95\% purity, supplier: Kurt J Lesker Company Ltd) and Si (99.999\% purity, supplier: Kurt J Lesker Company Ltd) targets in an argon atmosphere. The molybdenum was sputtered using a DC power supply in constant current mode and the silicon with an impedance matched radio frequency (RF) power supply. The depositions were carried out without intentional heating while rotating the substrate holder at a speed of 20 rotations per minute during deposition for better uniformity of the film growth. Prior to deposition, both targets were pre-sputtered at an argon pressure of $P$(Ar) = 2.5 mTorr for 5 minutes to stabilize the plasma and clean the surface. At first, the discharge current of the Mo target was fixed at 130 mA. The RF power applied to the Si target was then varied keeping other conditions unchanged in order to tune the composition of the MoSi films.
\begin{figure}
\includegraphics[width=.48\textwidth]{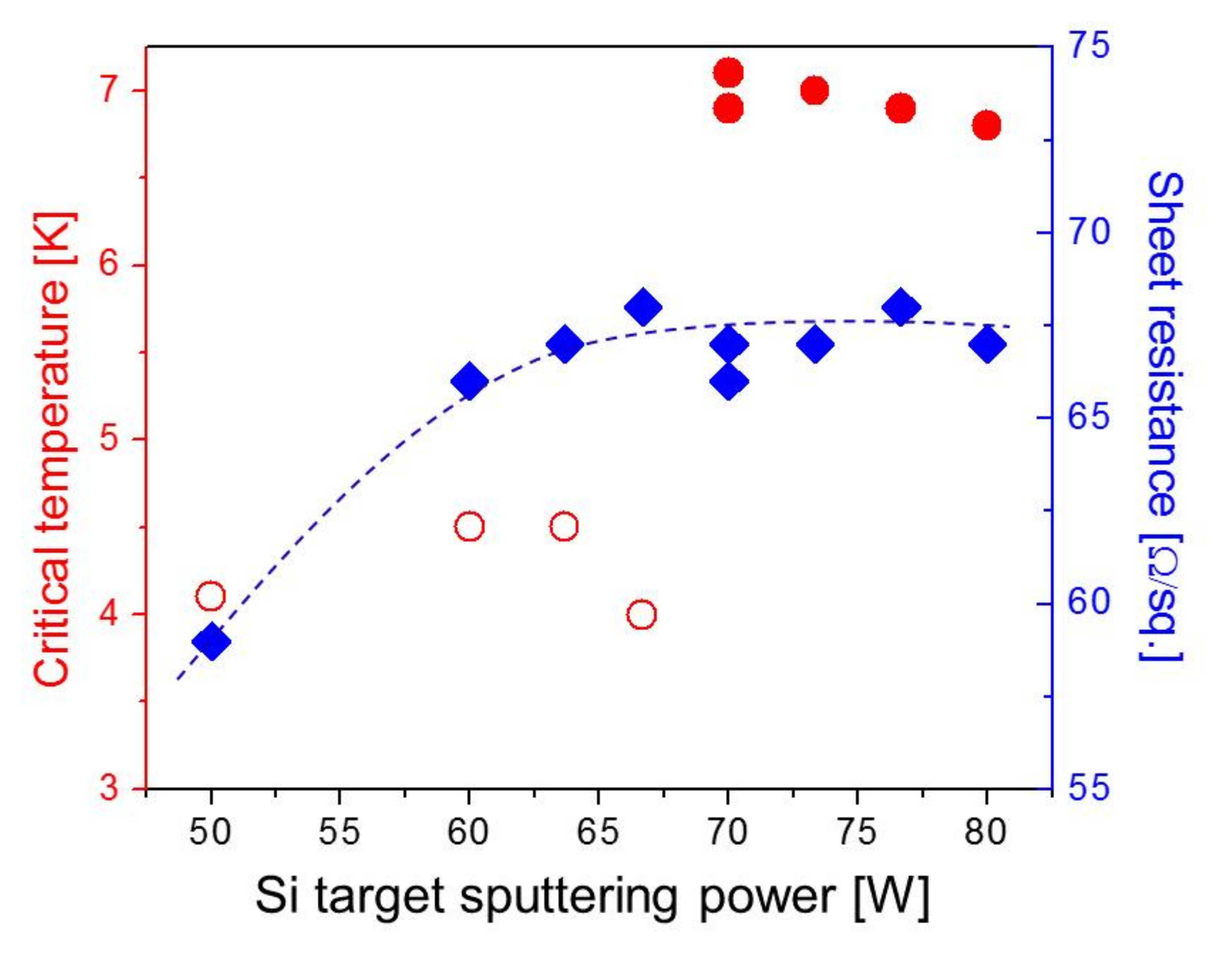}% Here is how to import EPS art
\caption{\label{fig:power}The dependence of critical temperature ($T_C$) (red dots) and the sheet resistance ($R_{\rm sh}$) (blue dots) on silicon content in 20~nm thick MoSi films. Open red dots show the upper limit for $T_C$ (i.e. the lowest measured  temperatures for samples where no superconducting transition was observed). The discharge current of the Mo target was fixed at 130 mA.}
\end{figure}

Fig.~\ref{fig:power} shows the critical temperature $T_C$ and sheet resistance $R_{\rm sh}$ of $\sim$20~nm thick MoSi films as a function of power to the Si target. The critical temperature is defined as the middle of the transition at $R$($T$) = $R_{\rm 20}$ / 2, where $R_{\rm 20}$ is a resistance at 20 K. The sheet resistance is measured by a four-probe measurement at room temperature. Above 70 W, the $T_C$ is weakly dependent ($\sim$7 K) on silicon content in the film (shown by filled red dots). Once the power was reduced below 70 W, no superconducting transition was observed at temperatures $\leq$ 4 K (the limit of our testing apparatus for $T_C$ screening). Open red symbols indicate the lowest temperatures measured for which the films remained normal (non-superconducting). This sharp transition is associated with a Mo rich composition reported in previous publications on MoSi film deposition\cite{Banerjee20177}; however, the transition to the lower $T_C$ (< 4 K) samples was found to be more gradual in their work than we observe here. The sheet resistance of films was found to be in the range 65-70 $\Omega$/sq. (blue dots in Fig.1) except one sputtered at the lowest power in the experiment (50 W) which was < 60 $\Omega$/sq.

To study the morphological structure of sputtered films we performed transmission electron dark-field imaging and diffraction on 7~nm thick MoSi film sputtered at conditions corresponding to the highest measured $T_C$ (73 W) using a field emission TEM (JEOL 2010F) at 200 kV. Additionally, we imaged Mo-rich film with the same thickness sputtered at 50 W of silicon-target power. 

\begin{figure}
\includegraphics[width=.48\textwidth]{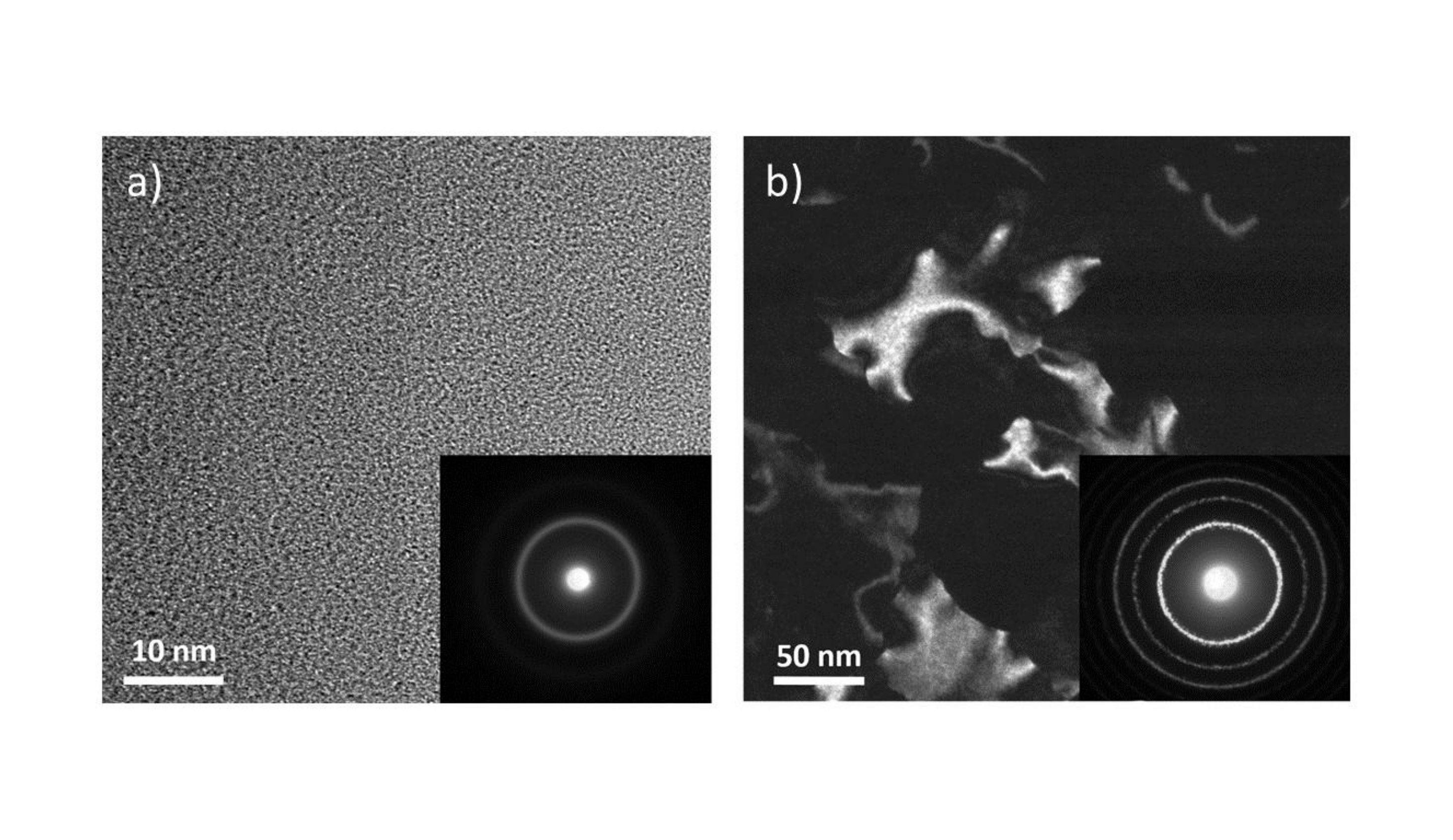}% Here is how to import EPS art
\caption{\label{fig:tem}Dark-field images of two 7~nm thick MoSi films taken by TEM. The images have different characteristic length scales. The insets show diffraction patterns. Films were sputtered at the following conditions: (a) 130 mA of DC current to the Mo target and 73 W of the RF power to the Si target. (b) 130 mA of the DC current to the Mo target and 50 W of the RF power to the Si target.}
\end{figure}

Results of examination by TEM are presented in Fig.~\ref{fig:tem}.
Fig.~\ref{fig:tem}(a) shows a dark-field image and diffraction pattern of the MoSi film sputtered at 73 W to the Si target.  The presence of a single diffuse ring in the diffraction pattern indicates that the film had limited crystallinity, which was confirmed by the dark-field TEM image. This image indicates that there were no big crystalline domains ($\sim$2~nm) in the film, in agreement with with x-ray diffraction (XRD) measurements implemented on 50-nm thick MoSi film. A detailed report of an investigation of the local structural ordering of amorphous superconducting MoSi films can be found in \cite{Banerjee20177}. Fig.~\ref{fig:tem}(b) shows a dark-field image and diffraction pattern of the MoSi film sputtered with 50 W applied to the Si target. The presence of sharp rings with distinct spots in the diffraction pattern indicates a polycrystalline structure with large grains. The dark-field image shows grains of size $\sim$50-100 nm. 

\begin{figure}
\includegraphics[width=.41\textwidth]{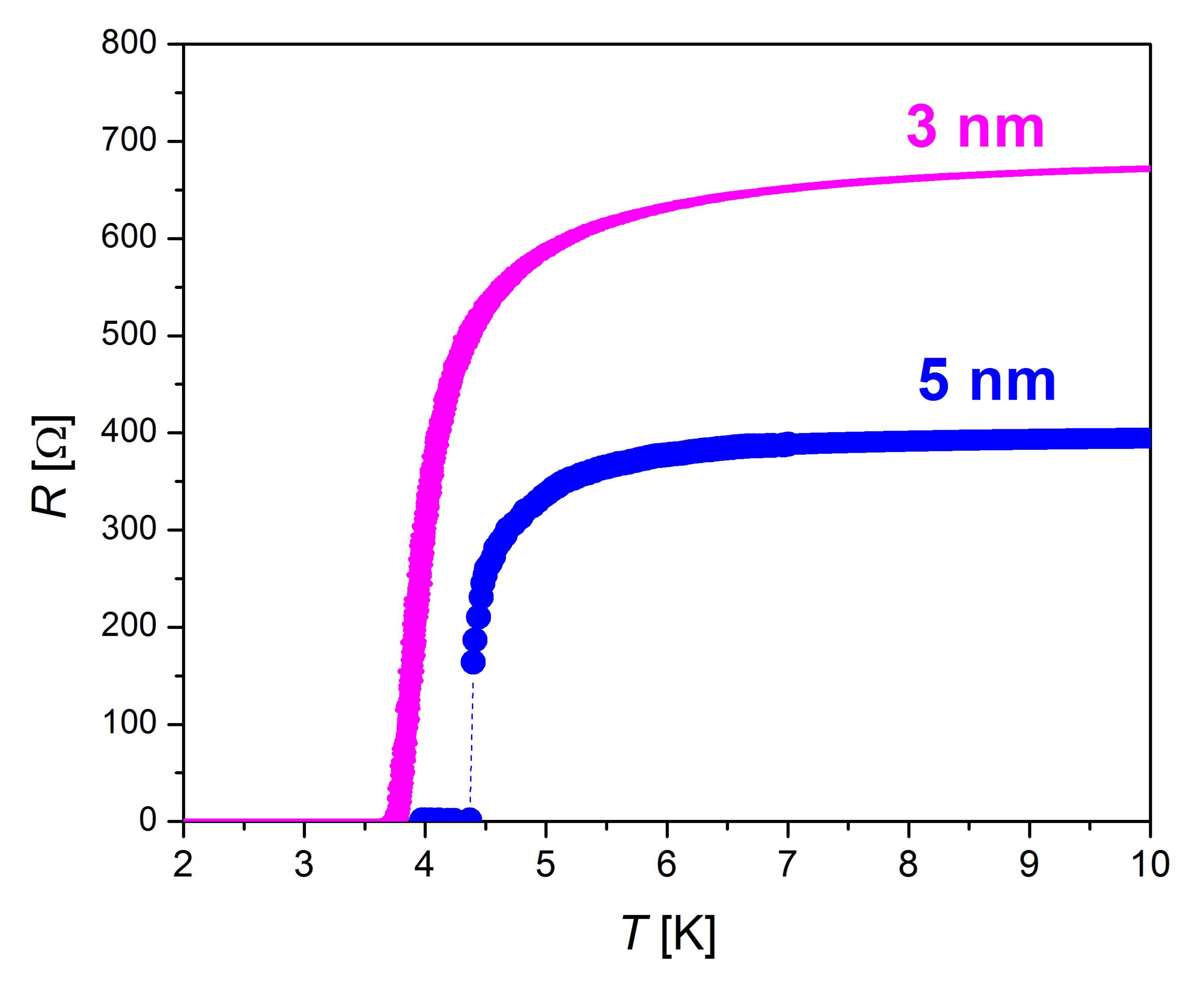}% Here is how to import EPS art
\caption{\label{fig:tc}$R$($T$)-dependence of superconducting MoSi films with thicknesses of 3 and 5 nm.}
\end{figure}

Based on the results of the film characterization, we decided to fabricate devices from MoSi film sputtered at 73 W. Superconducting MoSi films with thicknesses of 3 and 5 nm have been prepared for fabrication of single-photon detectors. Additionally, a 2~nm-thick Si layer was deposited \textit{in-situ}  on top of the MoSi films to prevent oxidation of the superconductor. Fig.~\ref{fig:tc} shows the resistance as function of the temperature plotted for both films. The critical temperature of the 5~nm film was found to be 4.6 K while the $T_C$ of the thinner film was around 4.1 K. Table~\ref{tab:table1} summarizes the electrical and superconducting properties of studied films and results reported in literature  on MoSi films for fabricated nanowires. While the critical temperature follows the trend of increase with an increase of the thickness, the width of transition was found to be the widest for our films.

We used electron-beam lithography with high-resolution positive electron-beam resist (ZEP 520 A) to pattern microwires. The resist was spin coated onto the chip at 5000 rpm which ensured a thickness of 335 nm. The resist was exposed to a 125 keV electron beam with an area dose density of 500 $\mu$C/cm$^2$. After exposure, the resist was developed by submerging the chip in O-xylene at 0 $^\circ$C for 90 s with subsequent rinsing in 2-propanol. The ZEP 520 A pattern was then transferred to the MoSi by reactive ion etching in CF$_4$ at 50 W for 6 minutes.

\begin{figure}
\includegraphics[width=.40\textwidth]{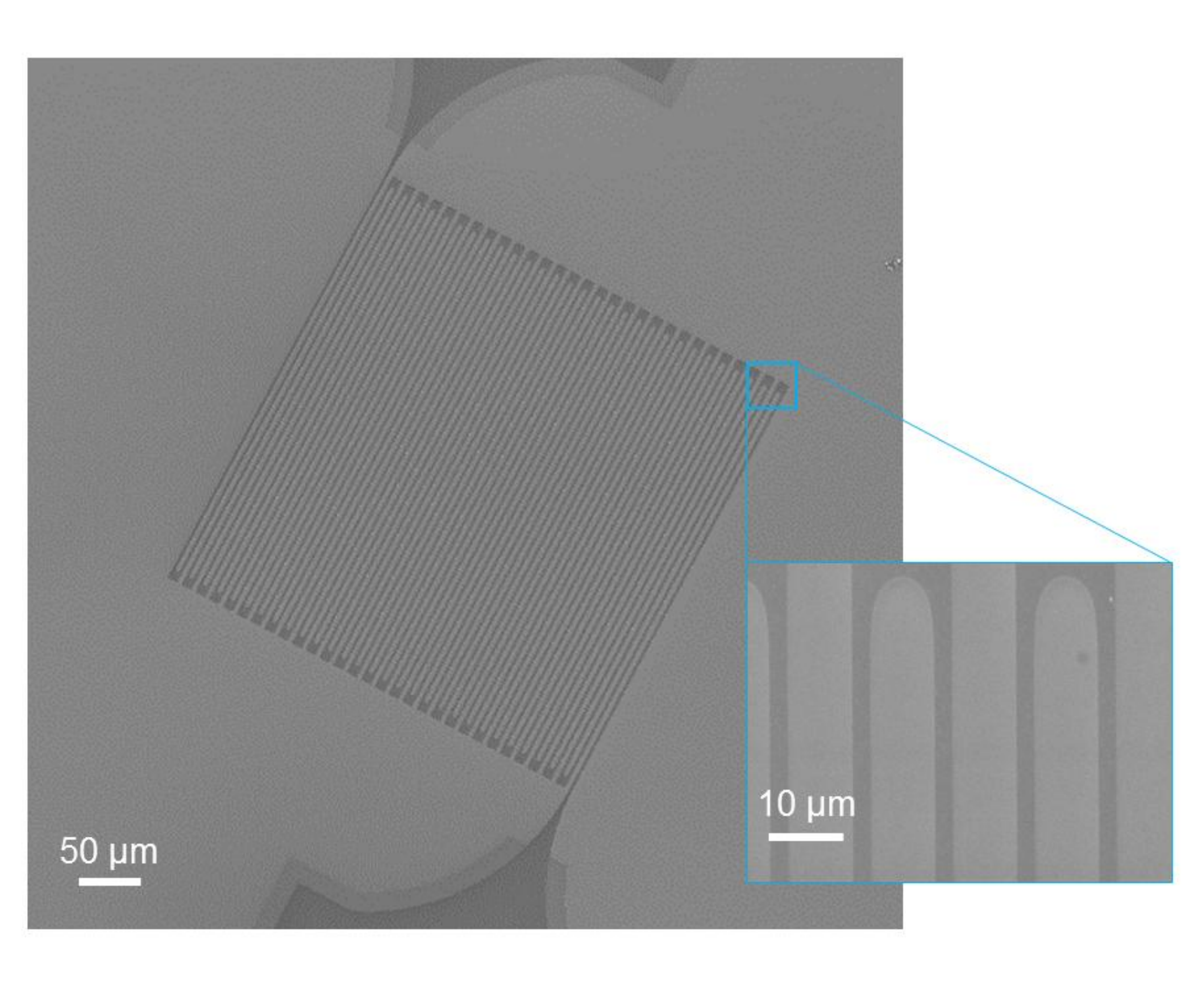}% Here is how to import EPS art
\caption{\label{fig:sem}SEM image of 3~$\mu$m microwire single-photon detector with surface area 400 by 400~$\mu$m$^2$ made of a 3~nm MoSi film.}
\end{figure}

Fig.~\ref{fig:sem} shows an SEM image of 3~$\mu$m microwire single-photon detector with surface area 400 by 400~$\mu$m$^2$. To minimize the current crowding\cite{Clem2011}, the fill factor was reduced to 23\%. In order to prevent latching \cite{Annunziata2010}, we designed microwires with a total number of squares  10$^4$ and 7$\cdot$10$^3$ for width of 1 and 3~$\mu$m respectively.
\begin{table}
\caption{\label{tab:table1}Superconducting and electrical properties of studied 3~nm and 5~nm thick MoSi films and reported characteristics of MoSi films for SNSPDs: $d$ - a thickness of a film; $R_{\rm 300}$ - a sheet resistance at 300 K; $RRR$ - a residual-resistance ratio; $T_C$ - a critical temperature; $\Delta$ $T$ - a width of transition.}
\begin{ruledtabular}
\begin{tabular}{ccddd}
$d$, nm&$R_{\rm 300}$, $\Omega$&\mbox{$RRR$}\footnote{ taken as $\Delta$ $T$=$T$(0.9$R_{\rm 20}$) - $T$(0.1$R_{\rm 20})$}&\mbox{$T_C$, K}&\mbox{$\Delta$ $T$, K}\\
\hline
3&585&\mbox{0.85}&\mbox{4.1}&\mbox{1.57}\\
5&345&\mbox{0.87}&\mbox{4.6}&\mbox{1.3}\\
2.8\cite{KorMoSi2018}&-&\mbox{-}&\mbox{3.1}&\mbox{-}\\
3.3\cite{Korneeva2020}&-&\mbox{-}&\mbox{3.85}&\mbox{-}\\
4\cite{Korneeva2014}&$\sim$700&\mbox{0.85-0.9}&\mbox{4.3-4.7}&\mbox{0.5}\\
5\cite{Caloz2018}&-&\mbox{0.94}&\mbox{4.9}&\mbox{1.1}\\
5\cite{Caloz2017}&-&\mbox{-}&\mbox{5}&\mbox{-}\\
5-10\cite{Banerjee20177}&200-450&\mbox{0.8-0.95}&\mbox{5.5-6.4}&\mbox{0.5-1}\\
10\cite{Li2016}&-&\mbox{-}&\mbox{6.5}&\mbox{0.4}\\
\end{tabular}
\end{ruledtabular}
\end{table} 

To electrically and optically characterize the fabricated device, we used an experimental setup in a single shot type He-3 cryostat with base temperature of 300~mK. The device was mounted on the sample holder using a contact glue (GE Varnish). The holder was screwed down to the cold stage. A low-temperature bias tee decoupled the high-frequency path from the DC bias path. The high-frequency signal was carried out of the cryostat by stainless-steel rigid coaxial cables, while DC bias was provided via a pair of twisted wires connected to a low-noise voltage source. The signal was amplified at room temperature by a low-noise amplifier LNA-2500 and then sent to a pulse counter. A single mode optical fiber guided light from a 1550~nm CW laser into the cryogenic apparatus through a vacuum feedthrough and was mounted on a stage above the sample surface. The single-photon counting regime was confirmed by measuring the count rate from the detectors, which was linear as a function of the applied laser power.
\begin{figure}
\includegraphics[width=.40\textwidth]{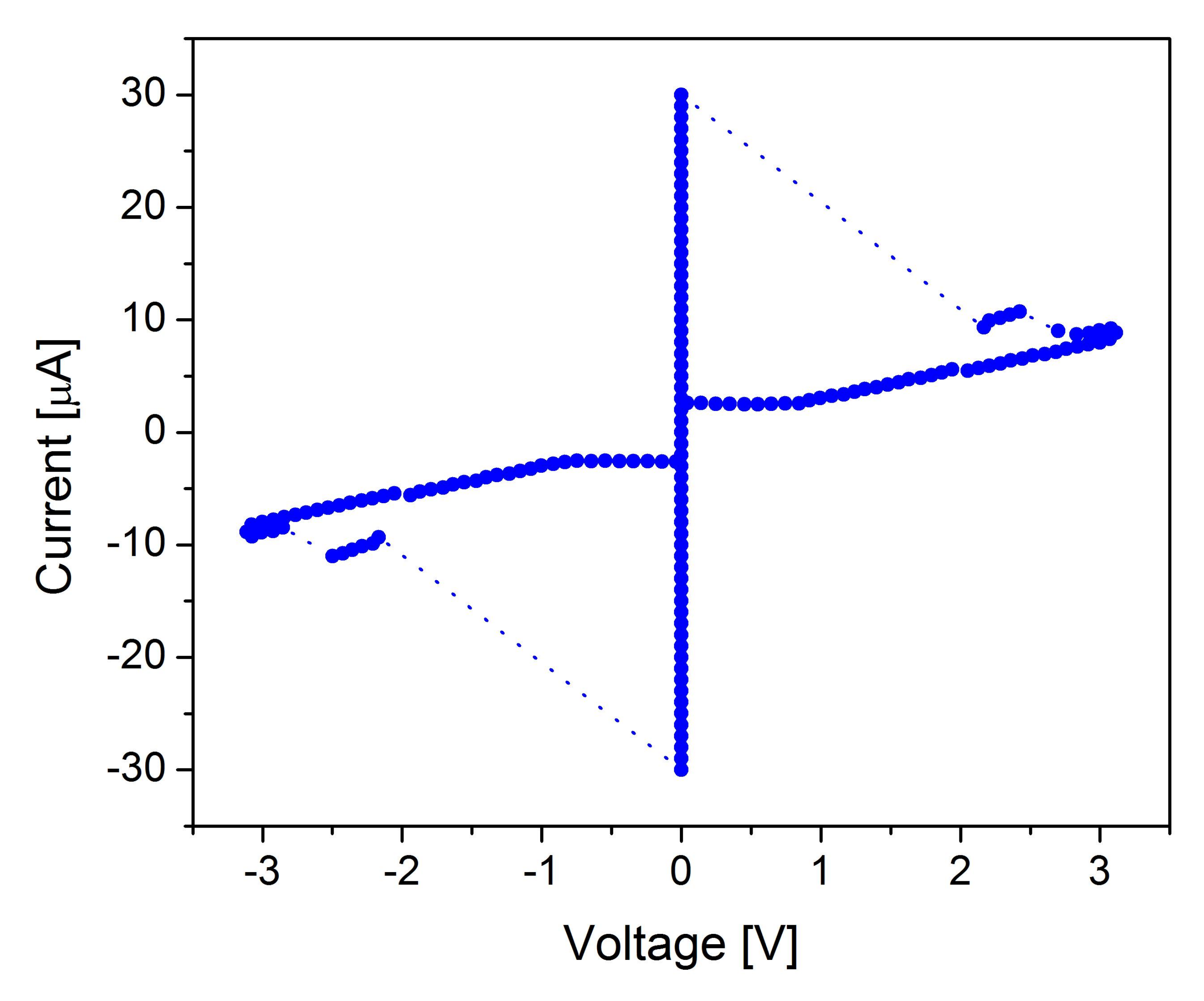}% Here is how to import EPS art
\caption{\label{fig:iv}An example of IV-trace for determining switching current.}
\end{figure}

We first characterized our devices in terms of their superconducting properties. Fig.~\ref{fig:iv} presents an example of an IV-curve measured in current-bias mode. For 3~nm thick devices, the switching current was found in the range of 8-10~$\mu$A and 27-30~$\mu$A for the 1 and 3~$\mu$m wide wires respectively. The measured switching current of 5~nm thick devices was 19-24~$\mu$A for 1~$\mu$m wires and 46-52~$\mu$A for 3~$\mu$m wires.
By extracting the switching current $I_{\rm SW}$ from IV-curves, we estimated the density of critical current in wires $j_C$ and then compared that with depairing critical current density $j_C^{\rm dep}$ \cite{Charaev2017}. To calculate $j_C^{\rm dep}$, we used a diffusion coefficient $D$ = 0.47 cm$^2$ s$^{\rm -1}$, experimentally obtained for MoSi film with similar thickness and critical temperature \cite{Korneeva2020}. We found the ratio to be in the range 0.5-0.63 for both 3 and 5~nm thick devices. 
\begin{figure*}
\includegraphics[width=.99\textwidth]{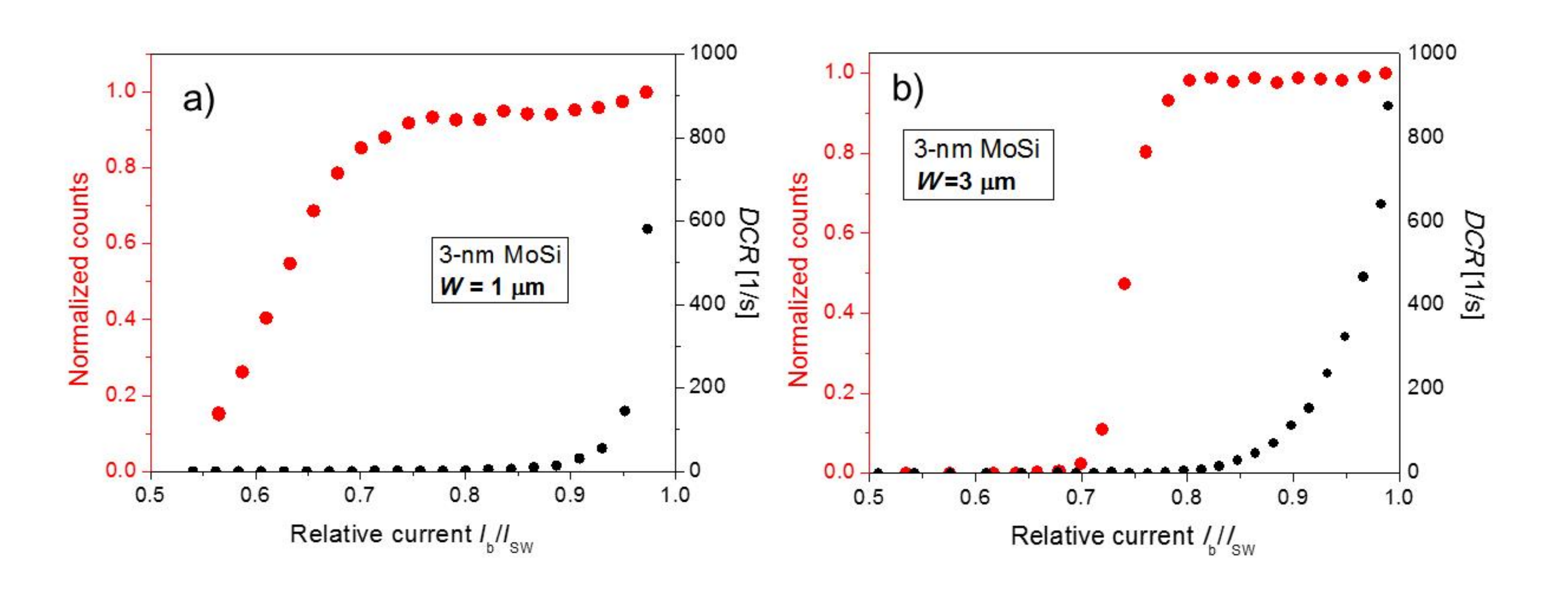}% Here is how to import EPS art
\caption{\label{fig:pcr}Normalized photon counts and dark count rate as a function of relative bias current taken at 300 mK from a) 1 $\mu$m and b) 3 $\mu$m wide detectors with thickness of 3 nm. 
Notice that both devices exhibited saturated count rate at high bias currents.}
\end{figure*}

We then performed optical measurements on both 3  and 5 nm chips. Fig.~\ref{fig:pcr}  represents photon count rate (PCR) and dark count rate (DCR) as a function of relative bias current for 3~nm thick MoSi detectors where $I_b$ is bias current. Both 1  and 3~$\mu$m devices show a clear plateau in the internal detection efficiency under 1550~nm illumination. The detection efficiency of the 3~nm thick devices with saturation is likely limited by the absorbance of the MoSi film. Using the optical properties of superconducting films\cite{Banerjee2018} we estimated the absorbance\cite{Semenov2009} of the 3~nm MoSi film to be 5.8\% at 1550 nm. 
The reset time of devices with lengths of 10$^4$ squares was determined to be 75 ns. From this value of the fall time and the geometry of the wide wires, we extracted a kinetic inductance of $\sim$375 pH/sq. for the 3 nm thick MoSi devices with width of 1 $\mu$m, corresponding to a penetration depth of $\lambda \sim 950\,{\rm nm}$. 

Despite $j_C$/ $j_C^{\rm dep}$ ratio of devices is similar on both chips, we did not observe optical response in devices of same nominal geometry made of 5 nm MoSi film.

The formation of a normal hot-spot with sufficient size after absorption of the photon is crucial for the emergence of a normal belt in wide wires. From a theoretical point of view\cite{Vod2017}, the radius of the hot-spot ($R_{\rm NS}$) is proportional to $\sqrt{E_{\rm ph} / N(0) d T_C^2}$,  where $N$(0) - the total density of electronic states at the Fermi level; $E_{\rm ph}$ - the energy of the photon. Assuming that the specific resistivity and diffusivity are similar for 3 and 5 nm MoSi films, the hot-spot radius size is depends on the critical temperature and the thickness for the same photon energy. The estimated $R_{\rm NS}$ for 5 nm MoSi is $\sim$ 50\% smaller than the hot-spot radius for thinner films. Taking into account the fact that the calculated depairing current is similar for samples of both thicknesses, single-photon sensitivity is expected in 5~nm MoSi wide devices for higher photon energies.

The absence of a photoresponse in 5~nm MoSi wide devices indicates possible issues in the realization of large-area wide detectors made of high-$T_C$ superconductors at telecom wavelength. Candidate materials for SNSPDs summarized in a recent overview\cite{Holzman2019} exhibited a broad range of critical temperature, resistivity, and optical properties. Obviously, low-$T_C$ materials are preferable due to the smaller superconducting energy gap. It also was shown that amorphous MoSi and sputtered polycrystalline TiN films show favorable optical properties at infrared wavelength\cite{Banerjee2018}. However, the number of internal constrictions in superconducting films should be minimized to reach the highest ratio of $j_C$/ $j_C^{\rm dep}$.

The dark-count rate was determined by blocking the optical path at the room-temperature end of fiber. All tested devices demonstrate DCR below 10$^3$ cps (examples are displayed on right axis in Fig.~\ref{fig:pcr}).

In conclusion, we fabricated and characterized 3 and 5~nm MoSi single-photon detectors with wire widths of 1 and 3 $\mu$m and surface area up to 400 by 400 $\mu$m$^2$ at 1550 nm wavelength. These detectors show saturated internal detection efficiency at a measurement temperature of 300 mK. The dark count rate was found to be two orders of magnitude lower than was measured on short devices with a shunt. Obtained results open new perspectives for scaling the detectors to masses large enough to probe new territory in the direct detection of sub-GeV dark matter by superconducting wires. Additionally, scaling the detector area up to the millimeter range allows collection of photons from resonant absorption targets \cite{Baryakhtar2018} with maximum efficiency and system-design flexibility.

The data that support the findings of this study are available from the corresponding author upon reasonable request.

\begin{acknowledgments}
The authors would like to thank J. Daley and M. Mondol of the MIT Nanostructures lab for the technical support related to electron-beam fabrication, and would like to thank Sae-Woo Nam and Jeff Chiles at NIST for helpful discussion and  sharing an early version of their manuscript with us, and also we would like to thank Dr. C. Settens from the Center for Materials Science and Engineering X-ray Facility for his assistance and advice on all matters related to x-ray measurements, and Brenden Butters and Mina Bionta for assistance in editing the final manuscript. Research was sponsored by the U.S. Army Research Office (ARO) and was accomplished under the Cooperative Agreement Number W911NF-16-2-0192.
\end{acknowledgments}

\bibliography{aipsamp}% Produces the bibliography via BibTeX.

\end{document}